# Ballistic pressure wave contributions to rapid incapacitation in the Strasbourg goat tests

Michael Courtney, PhD
Ballistics Testing Group, P.O. Box 24, West Point, NY 10996
Michael_Courtney@alum.mit.edu

Amy Courtney, PhD
Department of Physics, United States Military Academy, West Point, NY 10996
Amy_Courtney@post.harvard.edu

*Abstract:*
This article presents empirical models for the relationship between peak ballistic pressure wave magnitude and incapacitation times in the Strasbourg goat test data. Using a model with the expected limiting behavior at large and small pressure wave magnitudes, the average incapacitation times are highly correlated (R = 0.91) with peak pressure wave magnitude. The cumulative incapacitation probability as a function of time reveals both fast (t < 5 s) and slow (t > 5 s) incapacitation mechanisms. The fast incapacitation mechanism can be accurately modeled as a function of peak pressure wave magnitude. The slow incapacitation mechanism is presumably due to blood loss via damaged vascular tissue.
*Originally submitted 13 December 2006. Revised version submitted 1 August 2007.*

**I.      Introduction**
Selecting service caliber handgun loads with the greatest potential for rapid incapacitation of violent criminal or terrorist attackers is of interest to law enforcement [PAT89]. This interest has fueled heated debate on the merits of some contributions. The debate on whether a ballistic pressure wave plays a role in incapacitation has been reviewed elsewhere, and a hypothesis has been formed that a pressure wave can play a role in rapid incapacitation by handgun bullets [COC06a, COC06b].

**Pressure wave hypothesis:**
*Other factors being equal, bullets producing larger pressure waves incapacitate more rapidly than bullets producing smaller pressure waves.*

The pressure wave hypothesis received direct experimental support in *Experimental Observations of Incapacitation via Ballistic Pressure Wave Without a Wound Channel,* [COC07a], and traumatic brain injury has been linked to the ballistic pressure wave in *Links between traumatic brain injury and ballistic pressure waves originating in the thoracic cavity and extremities* [COC07b].

The Strasbourg tests [STR93] studied handgun bullet effectiveness in goats by shooting the test subjects broadside through the center of the chest and recording the time to incapacitation (falling down). These tests employed a pressure sensor inserted into the carotid artery of live unanaesthetized goats. These tests directly suggest that an internal pressure wave created by the interaction of the bullet and tissue can contribute to rapid incapacitation and can incapacitate more quickly than the crush cavity/blood loss mechanism alone:

*In a substantial number of cases, the subject was incapacitated almost instantly. Each time this occurred, between two and five pressure spike tracings of high amplitude and short duration were found which immediately preceded and matched corresponding, diffused, or flattened lines (EEG tracings). Normally, the time lag between the first pressure spike and the beginning of slowed or flattened lines was between 30 and 40 milliseconds (although there were several cases where this delay lasted as long as 80 milliseconds)…The taller pressure spike tracings always preceded the slowed or flat line tracing…The initial spikes had to be of a certain height in order for the animal to collapse immediately.*

The Strasbourg tests have been criticized on a number of counts [FAC94a, FAC97a]. A review of these criticisms [COC06a] shows them to be invalid and unconvincing because they contain numerous *ad hominem* attacks, appeal to unidentified authorities, and other fallacious reasoning. The review concludes:

*In the absence of support or direct contradiction from any other experiments, the veracity of the Strasbourg tests should fairly be considered to be an open question...Rather than lean too heavily on (possibly biased) expert opinions, the veracity of the report should be determined by the degree to which the reported results find support in other experimental findings.*

The relationship between pressure wave and incapacitation times in the Strasbourg tests are consistent with ballistic pressure waves causing remote brain injury in pigs [SHS90a, SHS90b, SHK90, SHL89, SHS88, SHS87], observations of remote brain injury in dogs [WWZ04], and observations of incapacitation and traumatic brain injury in fluid percussion model research [THG97, TLM05, and references therein]. The results also have quantitative agreement with an incapacitation study in deer [COC06d], and correlations between incapacitation and pressure wave magnitude in humans [COC06b].

Chamberlin observed damage remote from the wound channel he ascribed to the hydraulic reaction of body fluids [CHA66]. Tikka et al. showed that ballistic pressure waves originating in the thigh reach the abdomen. Wounding and delayed recovery of peripheral nerves have been reported [LDL45, PGM46]. Pressure waves cause compound action potentials in peripheral nerves [WES82], and ballistic pressure waves have been shown capable of breaking bones [MYR88].

Animal test subjects have been observed to be incapacitated by the ballistic pressure wave in the absence of a wound channel [COC07a], and numerous links between traumatic brain injury and the ballistic pressure wave have been documented [COC07b]. Remote ballistic pressure wave injury has also been reported in humans [OBW94]. Consequently, the principal observation of Strasbourg that pressure waves create incapacitation has considerable support in other research findings.

## II. Physics of the ballistic pressure wave

The origin of the pressure wave is Newton's third law. The bullet slows down in tissue due to the force the tissue applies to the bullet. By Newton's third law, the bullet exerts an equal and opposite force on the tissue. When a force is applied to a fluid or a visco-elastic material such as tissue or ballistic gelatin, a pressure wave radiates outward in all directions from the location where the force is applied.

The instantaneous magnitude of the force, $F$, between the bullet and the tissue is given by

$F = dE/dx$,

Where $E = \frac{1}{2} mV^2$ is the instantaneous kinetic energy of the bullet, and $x$ is the instantaneous penetration distance. $dE/dx$ is the first derivative of the energy with respect to the penetration depth. In other words, it is the local rate of kinetic energy loss per unit of penetration depth. Losing 100 ft-lbs of kinetic energy in 0.02 feet of penetration would create a force of 5,000 lbs because 100 ft-lbs/0.02 ft = 5,000 lbs.

This force (equal to the local rate of energy loss) changes continuously and depends on both the loss of velocity and the loss of mass (unless the mass is constant). By the chain rule of calculus,

$F = dE/dx = \frac{1}{2} V^2 \, dm/dx + m V \, dV/dx$,

where $dm/dx$ and $dV/dx$ are the local rates of mass and velocity loss with respect to penetration depth.

Applying this formula directly requires detailed knowledge of the instantaneous mass and velocity changes of a bullet at every point along the wound channel. The instantaneous force can be accurately estimated by shooting the same bullet through varying thicknesses of ballistic gelatin. In other words, one might shoot through a 0.05 ft thick block of gelatin to determine the loss of energy in the first 0.05 ft of penetration. Then one might shoot through a 0.1 ft thick block of gelatin to determine the loss of energy in a 0.1 ft thick block of gelatin. Then one might shoot through a 0.15 ft thick block of gelatin to determine the loss of energy in a 0.15 ft thick block of gelatin. Repeating this process using small increments, and applying standard techniques for estimating derivatives from measured values at closely spaced points would yield an accurate measurement of the instantaneous force at every penetration depth.

There are some simple and reasonable estimates that can be made more easily. In cases where the mass is constant, the average force $F_{ave}$ between the tissue and bullet is simply the initial kinetic energy $E$ divided by the penetration depth $d$.

$F_{ave} = E/d$.

For example, a bullet impacting with a kinetic energy of 500 ft-lbs and penetrating a depth of 12" (1 foot) exerts an average force of 500 lbs on the medium.

The peak of any variable force is larger than the average value. The peak force usually occurs during or soon after expansion, and most bullets have peak to average force ratios between 3 and 8. Bullets that do not expand, penetrate deeply, do not tumble (or tumble late) and lose energy gradually have a peak to average ratio close to 3. Bullets that expand rapidly, lose a lot of energy early, erode to a smaller diameter and then penetrate deeply have a peak to average ratio close to 8. Nosler Partition rifle bullets with their soft lead front section which expands rapidly and erodes away quickly leaving the base containing roughly 60% of the original mass at little more than the unexpanded diameter provide an example of large peak to average force ratio.

Most JHP handgun bullets have a peak to average ratio close to 5, so this can provide a reasonably accurate estimate of the peak force in many cases.

$F_{peak} = 5\, E/d$.

As discussed elsewhere [COC06b], fragmentation increases the peak retarding force.

Pressure is simply defined as force per unit area. This means that the pressure on the front of a bullet is simply the force divided by the frontal area of the bullet. The pressure exerted by the medium on the bullet is equal to the pressure exerted by the bullet on the medium. Because the frontal area of a bullet is small, the pressure at the front of the bullet is large.

Once created, this large pressure front travels outward from its source in all directions in a viscous or visco-elastic medium. As the wave propagates outward, the decrease in pressure magnitude is dominated by the increasing total area the pressure wave covers.

This decrease in pressure is analogous to the decrease in light intensity with increasing distance from a light bulb and the decrease in blast wave pressure with increasing distance from an explosion. The effective area of a pressure wave a distance R from the source is $A = 4\pi R^2$ (the surface area of a sphere of radius R). Consequently, the peak pressure P generated by an expanding bullet a distance R from the bullet path in a liquid or visco-elastic medium is

$$P = \frac{5E/d}{4\pi R^2}.$$

For example, the bullet impacting with a kinetic energy of 500 ft-lbs and penetrating to a depth of 1 foot creates a pressure wave with a peak magnitude of 796 PSI at the edge of a 1" diameter circle centered on the bullet path. We will use the symbol $P_{1"}$ to indicate the peak pressure at the edge of a 1" diameter circle centered on the bullet path.

The magnitude of a pressure wave will fall off with increasing distance from the point of origin unless reflected by a boundary or confined to an internal structure such as an artery. An internal pressure wave created in the thoracic cavity of an animal will be reflected many times by the sides of the cavity. The law of superposition can create localized regions of high pressure by focusing the wave, just as a concave mirror can focus a light wave and a concave surface can focus a sound wave.

In addition, a pressure wave confined to a tube will travel the length of the tube with little attenuation and can actually increase in magnitude if the tube narrows. This is analogous to a light wave confined to a fiber optic or the blast pressure wave from an explosion confined to a tunnel. Consequently, once the pressure wave reaches a major artery, it can be transmitted the length of the artery with only small attenuation. Think about hitting a fluid-filled garden hose with a hammer. The pressure wave created by the hammer strike will travel the length of the hose with relatively little loss of amplitude per foot of travel.

The fact that the average magnitude of the ballistic pressure wave is inversely proportional to penetration depth means that cutting the penetration depth in half (for a given amount of kinetic energy), doubles the pressure. However, this property will only increase incapacitation up to a point, because the pressure wave must be created inside of a visco-elastic medium and in close proximity to major blood vessels or vital organs to have its effect. A bullet that fails to penetrate into the thoracic cavity or that barely penetrates might have little effect. Once the penetration is below 9" or so, we expect the impact of the pressure wave will be reduced [COC06b].

The important parameter when considering blood loss as an incapacitation mechanism is the permanent crush cavity, which represents the crushed tissue that is left in the wake of a bullet. It can be estimated (FBI method) as the frontal area of the expanded bullet times the penetration depth. For predicting bullet effectiveness, this expanded area and penetration depth are commonly measured in 10% ballistic gelatin. The volume of the permanent crush cavity measured in this way is designated $V_{pcc}$ and is proportional to the penetration depth.

The fact that $V_{pcc}$ is linearly related to the penetration depth and that $P_{1"}$ is inversely related suggests that there might be a tradeoff between the two mechanisms because increasing penetration typically increases the crush volume while decreasing the pressure wave. Likewise, decreasing penetration increases the pressure wave but typically decreases the crush volume. Perhaps finding the "sweet spot" in this tradeoff is one key element in designing and selecting ammunition for a particular application and risk assessment. There has been much debate in the literature about the optimum penetration depth(s) for law enforcement and self-defense applications.

**III. Incapacitation Time Correlation with Bullet Performance Parameters**

The peak pressure wave magnitude (in PSI) on the edge of a 1" diameter cylinder concentric with the bullet path can be estimated as

$$P_{1"} = \frac{5E}{d\pi},$$

where E is the kinetic energy (in ft-lbs) of the bullet at impact, and d is the penetration depth (in feet).

One can consider what physical quantity related to bullet performance might be well correlated to the Strasbourg average incapacitation times. There is some difficulty attempting to correlate the average incapacitation times with bullet performance parameters of the full data set because many of these loads are no longer commercially available. Consequently, we consider a subset of loads for which there is sufficient data available in the published literature [MAS92, MAS96] to estimate the pressure wave and determine the bullet performance parameters of interest. (The loads we consider are listed in Appendix A.)

Performing linear least-squares fits (using a third order polynomial) to the average incapacitation times as a function of various performance parameters yields the correlation coefficients and standard errors shown in Table 1.

The average incapacitation times (AIT) were correlated with the following bullet parameters:

- $P_{1"}$: peak pressure wave magnitude determined on the surface of a 1" diameter cylinder centered on the wound channel.
- E: Kinetic energy of the bullet.
- TSC: Temporary stretch cavity volume.
- V: Bullet velocity.
- MV: Bullet momentum (mass times velocity).
- $V_{PCC}$: Permanent crush cavity volume estimated via FBI method.
- $V_{PCC}^{12"}$: Truncated permanent crush cavity volume estimated via FBI method truncated at 12" of penetration.
- $A_{PCC}$: Permanent crush cavity surface area estimated via FBI method.
- $A_{PCC}^{12"}$: Permanent surface area estimated via FBI method truncated at 12" of penetration.

These results show that AIT correlates most strongly with peak pressure wave magnitude, but also shows significant correlation with energy, velocity, and temporary cavity volume. Correlations with other bullet performance parameters are poor.

*Table 1: Correlation coefficients (R) and standard errors for AIT as a function of various bullet parameters. All these correlations used a linear least-squares fit to a third order polynomial.*

| Bullet Parameter | R | Standard Error (s) |
|---|---|---|
| $P_{1"}$ | 0.847 | 2.103 |
| E | 0.747 | 2.633 |
| TSC | 0.708 | 2.799 |
| V | 0.732 | 2.697 |
| MV | 0.512 | 3.402 |
| $V_{PCC}$ | 0.209 | 3.874 |
| $V_{PCC}^{12"}$ | 0.552 | 3.302 |
| $A_{PCC}$ | 0.316 | 3.758 |
| $A_{PCC}^{12"}$ | 0.319 | 3.744 |
| Initial diameter | 0.283 | 3.799 |

**IV. Model for average incapacitation time**

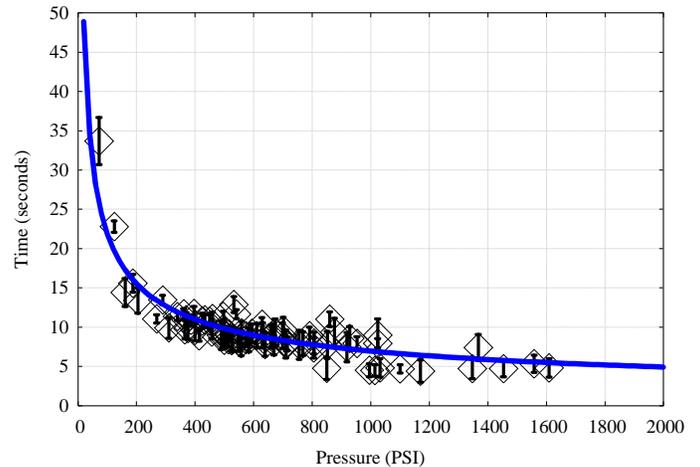

*Figure 1: A plot of average incapacitation time vs. pressure for the Strasbourg tests, along with the best-fit model.*

Developing a model with the proper limiting behavior for the average incapacitation time as a function of the peak pressure wave magnitude, p, yields:

$$AIT(p) = \frac{10s}{\sqrt{\dfrac{p}{p_o}}},$$

where $p_0$ is the characteristic pressure wave that gives an average incapacitation time of 10 seconds. Performing a least-squares fit gives $p_0$ = 482 PSI with a standard error of 1.64 s and a correlation coefficient of R = 0.91. A plot of AIT(p) is shown in Figure 1 along with the data.

The graph shows that while increasing the peak pressure wave magnitude from 100 PSI to 1000 PSI provides significant gains in reduced incapacitation times, increasing the peak pressure wave above 1000 PSI yields diminishing returns.

**V.       Cumulative probability distributions**

It is useful to graph the cumulative incapacitation probability distributions as a function of time. Because each load only has five data points, we group 10 different loads together for each cumulative probability distribution and use the average pressure wave to characterize the results. Group A combines the 10 loads with the lowest average incapacitation times. Group B combines the 10 loads with the next lowest 10 average incapacitation times, and so on.

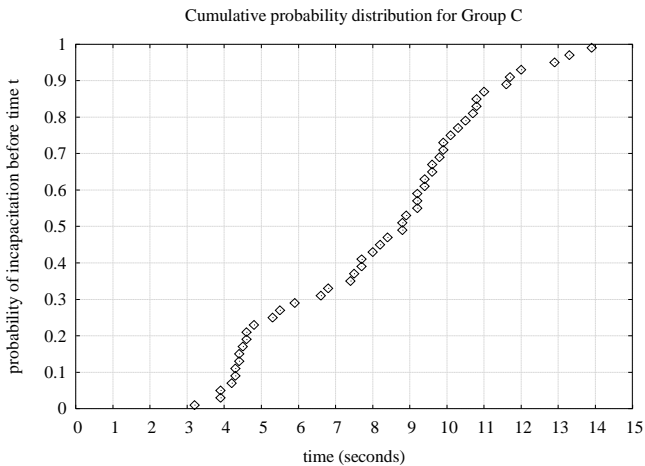

*Figure 2: Cumulative probability distribution for Group C.*

Figure 2 shows the cumulative probability distribution for Group C. These distributions all share the following characteristics: There is very little chance of creating truly immediate incapacitation. All distributions show 100% chance of eventual incapacitation. This is because a shot through the lungs will eventually cause a blood pressure drop sufficient for incapacitation. However, the different groups can be characterized by the degree to which they give evidence of rapid incapacitation (< 5 seconds).

Newgard made a compelling case that blood loss alone will very seldom produce incapacitation in under 5 seconds [NEW92]:

*For an average 70 kg (155 lb.) male the cardiac output will be 5.5 liters (~1.4 gallons) per minute. His blood volume will be 60 ml per kg (0.92 fl. oz. per lb.) or 4200 ml (~1.1 gallons). Assuming his cardiac output can double under stress (as his heart beats faster and with greater force), his aortic blood flow can reach 11 liters (~2.8 gallons) per minute. If one assumes a wound that totally severs the thoracic aorta, then it would take 4.6 seconds to lose 20% of his blood volume from one point of injury. This is the minimum time in which a person could lose 20% of his blood volume.*

These theoretical ideas are confirmed by many observations of deer almost always taking 5-10 seconds to fall with any broadside archery shot hit through the center of the chest. In contrast, we have observed numerous deer drop in under 5 seconds when hit by handgun bullets creating pressure waves at the larger end of the spectrum [COC06d]. Likewise, events of apparently involuntary incapacitation in under 5 seconds are repeatedly reported in humans for handgun shots which fail to hit the CNS or supporting bone structure.

Therefore, we will model the cumulative probability distribution, P(t), as having a fast and a slow component.

$$P(t) = \left[1 - P_{fast}^{fail}(t) P_{slow}^{fail}(t)\right],$$

where $P_{fast}^{fail}(t)$ is the probability of the fast mechanism failing to create incapacitation after a time t, and $P_{slow}^{fail}(t)$ is the probability of the slow mechanism failing to cause incapacitation after time t. Combining these independent probabilities using the product rule, and using the complementary rule to compute the probability for successful incapacitation after time t gives the equation above for P(t).

To model P(t), we model the slow probability of failure to incapacitate as:

$$P_{slow}^{fail}(t) = e^{-(t/t_s)^6}.$$

Note that this function has the expected limiting behavior of failing to incapacitate 100% of the time at t = 0, and failing 0% of the time for very large times. The characteristic time scale of the slow incapacitation mechanism is $t_s$, which presumably depends on the rate of blood loss.

To model P(t), we model the fast probability of failure to incapacitate as:

$$P_{fast}^{fail}(t) = \frac{A_0}{1+\left(\dfrac{t}{t_f}\right)^4} + 1 - A_0.$$

Note that this function has the expected limiting behavior of failing to incapacitate 100% of the time at t = 0, and succeeding to incapacitate $A_0$ of the time for very large times. The characteristic time scale of the fast incapacitation mechanism is $t_f$. $A_0$ describes the eventual incapacitation probability associated with the fast mechanism. Note that, unlike the blood loss mechanism, the probability of eventual incapacitation with the fast mechanism is less than 100%. After fitting the Strasbourg data to determine $t_f$ and $A_0$ for several groups with different pressure magnitudes, we will show that $t_f$ and $A_0$ are correlated with peak pressure wave magnitude.

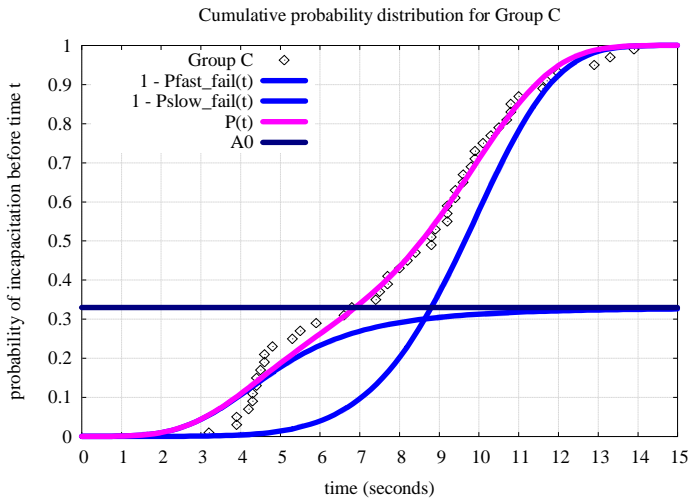

*Figure 3: Cumulative probability distribution of Group C shown along with best fit model P(t) in pink, component probability models for the fast and slow mechanisms in blue, and A0, the maximum probability of the fast mechanism eventually working, shown in black.*

Performing a least-squares fit of P(t) to determine the parameter values $A_0$, $t_f$, and $t_s$ that produce the closest correspondence with the experimental probability distribution for Group C yields: $A_0$ = 0.329(24), $t_f$ = 4.82 s (.23), and $t_s$ = 10.2507 s (.086).[1] Figure 3 shows the best-fit P(t) along with the component slow and fast probability curves for the best fit values of Group C. This model and fitting results can be interpreted to imply that a load with a peak pressure of 683 PSI will produce rapid incapacitation roughly 33% of the time with a typical time scale of 4.8 s.

*Table 2: Pressure wave magnitude, p; eventual fast incapacitation probability, $A_0$; fast incapacitation characteristic time, $t_f$; and slow incapacitation characteristic time, $t_s$ for Group A through Group F. (Uncertainty in last significant digit(s) is shown in parentheses.)*

| Group | P (PSI) | $A_0$ | $t_f$ (s) | $t_s$ (s) |
|---|---|---|---|---|
| A | 1221(86) | 0.645(13) | 2.94(5) | 8.18(11) |
| B | 829(60) | 0.433(16) | 4.14(12) | 10.73(9) |
| C | 683(65) | 0.329(24) | 4.82(23) | 10.25(9) |
| D | 611(42) | 0.195(29) | 5.13(54) | 10.07(8) |
| E | 650(53) | 0.393(15) | 5.36(16) | 12.13(8) |
| F | 554(57) | 0.379(4) | 6.56(39) | 12.33(14) |

Repeating this process for Group A through Group F[2] produces the values of $A_0$, $t_f$, and $t_s$ shown in Table 2, along with the average peak pressure magnitude of these groups. There is a trend that the maximum probability of the fast mechanism producing incapacitation ($A_0$), increases with pressure wave magnitude. This trend is unclear below 700 PSI because of the relatively small probability of the fast incapacitation mechanism at those low pressure levels. However, the trend is clear above 700 PSI. Likewise, there is a trend that the time scale of the fast mechanism, $t_f$, tends to decrease with increasing pressure wave magnitude.

There is a clearer trend that the time scale of the fast mechanism, $t_f$, tends to decrease with increasing pressure wave magnitude. The fast response time, $t_f$, is inversely proportional to the peak pressure wave magnitude, as shown in Figure 5. The solid curve is the result of a least-squares fit to the function

$$t_f(p) = \frac{10s}{p/p_f},$$

where $p_f$ = 342(8) PSI gives a correlation coefficient of R = 0.92 with the data of fast response time vs. pressure wave magnitude. This shows that the fast response time can be well described in terms of the pressure wave.

---

[1] Numbers in parentheses represent estimated uncertainty in least significant digits.

[2] At incapacitation times/pressure wave magnitudes smaller than Group F, the fast response mechanism is no longer discernable in the probability curves. Since only the slow mechanism gives clear evidence of acting in Groups G and H, these data sets cannot be used to reliably estimate $A_0$ and $t_f$ for the smaller pressure wave magnitudes.

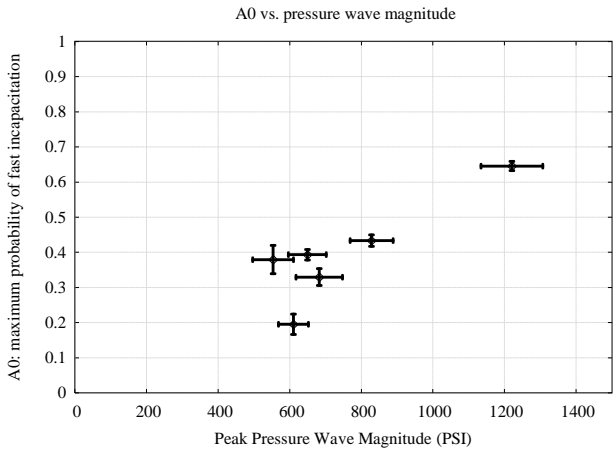

*Figure 4: Probability of eventual incapacitation via the fast mechanism plotted vs. pressure wave magnitude for Group A through Group F. There is no clear trend below 700 PSI, but the trend is clearly increasing above 700 PSI.*

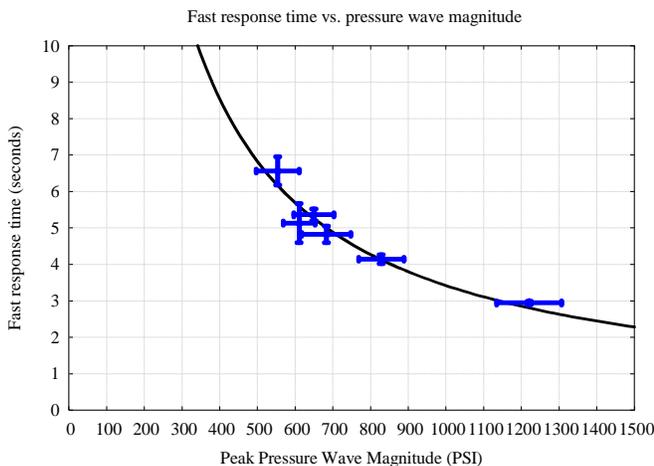

*Figure 5: Fast response time vs. pressure wave magnitude for Group A through Group F. There is a clear trend that the fast response time is inversely proportional to the peak pressure wave magnitude.*

## VI. Discussion

This analysis has shown that the Strasbourg test incapacitation times are highly correlated to peak pressure wave magnitude. Both the average incapacitation times and the probability distribution of incapacitation times demonstrate pressure wave effects. There appears to be both fast and slow mechanisms in play. The fast mechanism depends strongly on peak pressure wave magnitude and is easily discernable and fast acting (under 5 seconds) for pressure wave magnitudes above 700 PSI on the diameter of a 1" diameter cylinder concentric with the wound channel.

This shows that, all other factors being equal, bullets that produce pressure waves of greater magnitude incapacitate more rapidly than bullets that produce smaller pressure waves. The Strasbourg test data convincingly supports the pressure wave hypothesis and allows (perhaps for the first time) the fast response time to be modeled as a function of peak pressure wave magnitude.

It is worth noting that the $A_0$ determined from Group A is too small to model accurately[3] with a one parameter S curve increasing monotonically from 0 to 1 with increasing pressure wave magnitude. This is consistent with observations in deer. Even at rifle levels of peak pressure wave magnitude (p > 2000 PSI), a significant fraction of deer remain on their feet for over 5 seconds. The goat model for $A_0$ shown in Figure 6 increases monotonically from 0 to 0.7 to account for these observations.

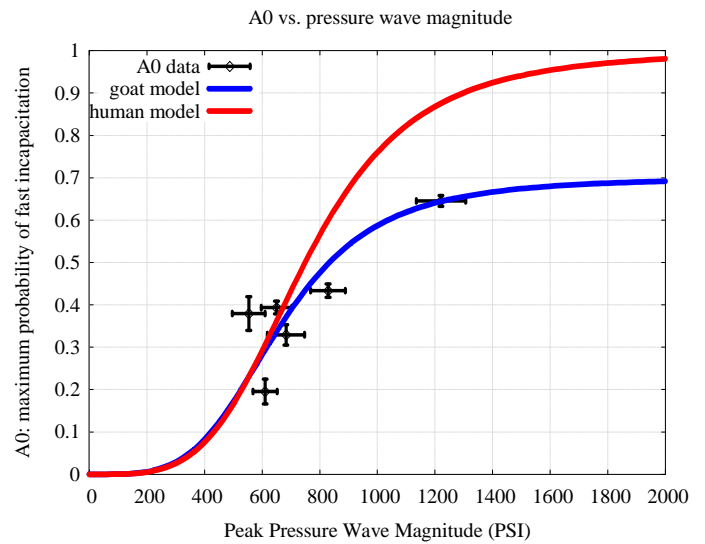

*Figure 6: Possible models for $A_0(p)$ in goats and humans compared with goat data.*

On the other hand, Marshall and Sanow [MAS96] observe that nearly all humans are instantly incapacitated by a shot to the chest with a .308 using expanding or fragmenting bullets (that create a very large pressure wave). This suggests that humans are more sensitive to pressure wave effects than deer and goats, with the asymptotic behavior of $A_0$ closer to 1. Behavior of $A_0$ in humans is probably more like the red curve in Figure 6 than the blue curve.

---

[3] *One might think that the large error bars in the data set would preclude a high level of confidence in this statement. However, we have looked at the data carefully with a number of different exploratory data analysis techniques, and we are confident that $A_0$ does not approach 1 asymptotically.*

One might note a seeming contradiction of an average incapacitation time that is inversely proportional to $p^{1/2}$ and a fast incapacitation time $t_f$ that is inversely proportional to $p$. There is no contradiction, because the data set as a whole (and thus the general trend of the average incapacitation time in Figure 1) depends more strongly on the slow incapacitation mechanism, which is presumably dominated by blood loss caused by vascular tissue damage. Both the cross sectional area and surface area of damaged tissue are expected to be roughly proportional to $p^{1/2}$ if we consider a region of tissue in which the pressure wave magnitude exceeds some pressure threshold required for vascular damage. In contrast, the pressure magnitude reaching the central nervous system is expected to be proportional to $p$.

Even if the data set (as graphed in Figure 1) extended into sufficiently high pressure ranges for the fast mechanism to dominate, the 30% of cases where the fast mechanism did not occur would still determine the scaling of average incapacitation times with pressure wave magnitude, and it would remain inversely proportional to $p^{1/2}$ in goats and deer. In contrast, the scaling of average incapacitation times in humans would probably be inversely proportional to $p$ for large $p$. As $A_0$ approaches 1, the fast mechanism completely dominates the average incapacitation time.

*A.        Cautions and Limits of Interpretation*
Goats are not people, and the shot angle used in the Strasbourg tests was particularly favorable to loads with shallow penetration. It would be an error to infer that the loads that worked well in the goat tests would necessarily work well in self-defense applications with a variety of shot angles and different penetration requirements.

Do not be overly impressed by the propensity for shallow penetrating loads to produce larger pressure waves. Bullet selection criteria should first determine the required penetration depth for the given risk assessment and application, and only use pressure wave magnitude as a selection criterion for bullets which meet a minimum penetration requirement.

Reliable expansion, penetration, feeding, and functioning are all important aspects of load testing and selection. It would be unwise to abandon long-held criteria of the load testing and selection process, but it seems prudent to consider the pressure wave magnitude along with other factors.

*B.        Implications for Bullet Design*
The trend in bullet design over the last decade has drifted toward bullets with little fragmentation and a higher percentage of retained mass. Bullets that both fragment and meet minimum penetration requirements create larger pressure wave magnitudes and offer improved incapacitation potential [COC06b].

In addition to moving toward designs which both penetrate and fragment reliably, the incapacitation potential of a bullet can be further improved by delaying expansion and fragmentation to a penetration depth of at least 4". This would place the peak pressure magnitude closer to vital organs.

Optimal use of a bullet's kinetic energy to produce pressure wave incapacitation suggests a bullet design that penetrates the first 4" or so prior to significant expansion or energy loss, and then rapidly expands and transfers a large percentage of its energy and 40% of its mass at penetration depths between 4-8" before continuing to penetrate to the depth desired for the application.

**REFERENCES/BIBLIOGRAPHY:**

[CHA66] Chamberlin FT, Gun Shot Wounds, in Handbook for Shooters and Reloaders, Vol. II, Ackley PO, ed., Plaza Publishing, Salt Lake City, Utah, 1966.

[COC06a] Courtney M, Courtney A: Review of criticisms of ballistic pressure wave experiments, the Strasbourg goat tests, and the Marshall and Sanow data, 2006.
http://arxiv.org/ftp/physics/papers/0701/0701268.pdf

[COC06b] Courtney M, Courtney A: Relative incapacitation contributions of pressure wave and wound channel in the Marshall and Sanow data set, 2006.
http://arxiv.org/ftp/physics/papers/0701/0701266.pdf

[COC06c] Courtney M, Courtney A: Ballistic pressure wave contributions to rapid incapacitation in the Strasbourg goat tests, 2006.
http://arxiv.org/ftp/physics/papers/0701/0701267.pdf

[COC06d] Courtney M, Courtney A: A method for testing handgun bullets in deer, 2006.
http://arxiv.org/ftp/physics/papers/0702/0702107.pdf

[COC06e] Courtney M and Courtney A: Using sound of target impact for acoustic reconstruction of shooting events, submitted to AFTE Journal, 2006.
http://www.ballisticstestinggroup.org/impactsound.pdf

[COC07a] Courtney M, Courtney A: Experimental Observations of Incapacitation via Ballistic Pressure Wave without a Wound Channel, 2007.
http://www.ballisticstestinggroup.org/lotor.pdf

[COC07b] Courtney A, Courtney M: Links between traumatic brain injury and ballistic pressure waves originating in the thoracic cavity and extremities. Brain Injury 21(7): 657-662, 2007. Pre-print:
http://www.ballisticstestinggroup.org/tbipwave.pdf

[EWL98] Evan AP, Willis LR Lingeman JE, McAteer JA: Editorial: Renal Trauma and the Risk of Long-Term Complications in Shock Wave Lithotripsy. Nephron 78(1):1-8, 1998.

[FAC88a] Fackler ML: Wound Ballistics: A review of common misconceptions. Journal of the American Medical Association 259:2730-2736; 1988.

[FAC94a] Fackler ML: The 'Strasbourg Tests:' Another Gunwriter/Bullet Salesman Fraud? Wound Ballistics Review 1(4):10-11; 1994.

[FAC96a] Fackler ML: Gunshot Wound Review. Annals of Emergency Medicine 28(1): 194-203; 1996.

[FAC97a] Fackler ML: Book Review: Street Stoppers – The Latest Handgun Stopping Power Street Results. Wound Ballistics Review 3(1):26-31; 1997.

[FAC99a] Fackler ML: Editorial. Wound Ballistics Review 4(2):15-16; 1999.

[FBC89] Fackler ML, Breteau JPL, Courbil LJ, et al.: Open Wound Drainage Versus Wound Excision in Treating the Modern Assault Rifle Wound. Surgery 105(5):576-84; 1989.

[GIK88] Göransson AM, Ingvar DH, Kutyna F: Remote Cerebral Effects on EEG in High-Energy Missile Trauma. The Journal of Trauma. 28(1 Supplement):S204-S205; January 1988.

[HKO47] Harvey EN, Korr IM, Oster G, et al.:Secondary damage in wounding due to pressure changes accompanying the passage of high velocity missiles. Surgery 21:218-239; 1947.

[KNO03] Knudsen SK, Oen EO: Blast-induced Neurotrauma in Whales. Neuroscience Research 46(3):377-386; 2003.

[LAG16] LaGarde LA: Gunshot Injuries ed 2. New York: William Wood & Company, 1916.

[LDL45] Livingstone WK, Davis EW, Livingstone KE: Delayed recovery in peripheral nerve lesions caused by high velocity wounding. J. Neurosurg., 2: 170, 1945.

[LKK03] Lingeman JE, Kim SC, Keo RL, McAteer JA, Evan AP: Shockwave Lithotripsy: Anecdotes and Insights. Journal of Endourology 17(9):687-693; 2003.

[LOS01] Lokhandwalla M, Sturtevant B: Mechanical Haemolysis in Shock Wave Lithotripsy (SWL): I. Analysis of Cell Deformation due to SWL Flow-Fields." Physics in Medicine & Biology 46(2):413-437; 2001.

[MAC94] MacPherson D: Bullet Penetration—Modeling the Dynamics and the Incapacitation Resulting From Wound Trauma. Ballistics Publications, El Segundo, CA, 1994.

[MAS92] Marshall EP and Sanow EJ: Handgun Stopping Power: The Definitive Study. Paladin Press, Boulder, CO, 1992.

[MAS96] Marshall EP and Sanow EJ: Street Stoppers. Paladin Press, Boulder, CO, 1996.

[MAS01] Marshall EP and Sanow EJ: Stopping Power. Paladin, Boulder, CO, 2001.

[MYR88] Ming L, Yu-Yuan M, Ring-Xiang F, Tian-Shun F: The characteristics of pressure waves generated in the soft target by impact and its contribution to indirect bone fractures. The Journal of Trauma 28(1) Supplement: S104-S109; 1988.

[NEW92] Newgard, Ken, MD: The Physiological Effects of Handgun Bullets: The Mechanisms of Wounding and Incapacitation. Wound Ballistics Review, 1(3): 12-17; 1992.

[OBW94] Ordog GJ, Balasubramanian S, Wasserberger J, et al.: Extremity Gunshot Wounds. I. Identification and Treatment of Patients at High Risk of Vascular Injury. The Journal of Trauma 36:358-368; 1994.

[PAT89] Patrick UW: Handgun Wounding Factors and Effectiveness. FBI Firearms Training Unit, Quantico, VA. 1989.

[PGM46] Puckett WO, Grundfest H, McElroy WD, McMillen JH, Damage to peripheral nerves by high velocity missiles without a direct hit. J. Neurosurg., 3: 294, 1946.

[ROW92] Roberts GK, Wolberg EJ: Book Review: Handgun Stopping Power: The Definitive Study. AFTE Journal 24(4):10; 1992.

[SAN96] Sanow E: Predicting Stopping Power. Handguns November; 1996.

[SBC01] Sokolov DL, Bailey MR, Crum LA: Use of a Dual-Pulse Lithotriptor to Generate a Localized and Intensified Cavitation Field. Journal of the Acoustical Society of America 110(3):1685-1695, 2001.

[SHA02] Shaw NA: The Neurophysiology of Concussion. Progress in Neurobiology 67:281-344; 2002.

[SHK90] Suneson A, Hansson HA, Kjellström BT, Lycke E, and Seeman T: Pressure Waves by High Energy Missile Impair Respiration of Cultured Dorsal Root Ganglion Cells. The Journal of Trauma 30(4):484-488; 1990.

[SHL89] Suneson A, Hansson HA, Lycke E: Pressure Wave Injuries to Rat Dorsal Cell Ganglion Root Cells in Culture Caused by High Energy Missiles, The Journal of Trauma. 29(1):10-18; 1989.

[SHS87] Suneson A, Hansson HA, Seeman T: Peripheral High-Energy Missile Hits Cause Pressure Changes and Damage to the Nervous System: Experimental Studies on Pigs. The Journal of Trauma. 27(7):782-789; 1987.

[SHS88] Suneson A, Hansson HA, Seeman T: Central and Peripheral Nervous Damage Following High-Energy Missile Wounds in the Thigh. The Journal of Trauma. 28(1 Supplement):S197-S203; January 1988.

[SHS90a] Suneson A, Hansson HA, Seeman T: Pressure Wave Injuries to the Nervous System Caused by High Energy Missile Extremity Impact: Part I. Local and Distant Effects on the Peripheral Nervous System. A Light and Electron Microscopic Study on Pigs. The Journal of Trauma. 30(3):281-294; 1990.

[SHS90b] Suneson A, Hansson HA, Seeman T: Pressure Wave Injuries to the Nervous System Caused by High Energy Missile extremity Impact: Part II. Distant Effects on the Central Nervous System. A Light and Electron Microscopic Study on Pigs. The Journal of Trauma. 30(3):295-306; 1990.

[STR93] The Strasbourg Tests, presented at the 1993 ASLET International Training Conference, Reno, Nevada.

[TCR82] Tikka S, Cederberg A, Rokkanen P: Remote effects of pressure waves in missile trauma: the intra-abdominal pressure changes in anaesthetized pigs wounded in one thigh. Acta Chir. Scand. Suppl. 508: 167-173, 1982.

[THG97] Toth Z, Hollrigel G, Gorcs T, and Soltesz I: Instantaneous Perturbation of Dentate Interneuronal Networks by a Pressure Wave Transient Delivered to the Neocortex. The Journal of Neuroscience 17(7);8106-8117; 1997.

[TLM05] Thompson HJ, Lifshitz J, Marklund N, Grady MS, Graham DI, Hovda DA, McIntosh TK: Lateral Fluid Percussion Brain Injury: A 15-Year Review and Evaluation. Journal of Neurotrauma 22(1):42-75; 2005.

[WES82] Wehner HD, Sellier K: Compound action potentials in the peripheral nerve induced by shockwaves. Acta Chir. Scand. Suppl. 508: 179, 1982.

[WWZ04] Wang Q, Wang Z, Zhu P, Jiang J: Alterations of the Myelin Basic Protein and Ultrastructure in the Limbic System and the Early Stage of Trauma-Related Stress Disorder in Dogs. The Journal of Trauma. 56(3):604-610; 2004.

[WOL91] Wolberg EJ: Performance of the Winchester 9mm 147 Grain Subsonic Jacketed Hollow Point Bullet in Human Tissue and Tissue Simulant. Wound Ballistics Review. Winter 91:10-13; 1991.

**About the Authors**

*Amy Courtney* currently serves on the faculty of the United States Military Academy at West Point. She earned a MS in Biomedical Engineering from Harvard University and a PhD in Medical Engineering and Medical Physics from a joint Harvard/MIT program. She has taught Anatomy and Physiology as well as Physics. She has served as a research scientist at the Cleveland Clinic and Western Carolina University, as well as on the Biomedical Engineering faculty of The Ohio State University.

*Michael Courtney* earned a PhD in experimental Physics from the Massachusetts Institute of Technology. He has served as the Director of the Forensic Science Program at Western Carolina University and also been a Physics Professor, teaching Physics, Statistics, and Forensic Science. Michael and his wife, Amy, founded the Ballistics Testing Group in 2001 to study incapacitation ballistics and the reconstruction of shooting events. www.ballisticstestinggroup.org

Revision information:
*13 December 2006 to 1 August 2007:* Fixed typographical errors. Updated references, contact information, and biographies. Added references [CHA66], [LDL45], [PGM46], [MYR88], [WES82], [TCR82], [COC07a] and [COC07b]. Table 1 adjusted slightly for corrected $A_{PCC}^{12"}$ values because one load had this parameter computed incorrectly.

# Appendix A: Data Table

| Cartridge | Load | V (fps) | MV (lbs-sec) | $A_{PCC}$ (sq. in.) | $A_{PCC}^{12"}$ (sq. in.) | $V_{PCC}$ (cu. in.) | $V_{PCC}^{12"}$ (cu. in.) | TSC (cu. in.) | E (ft-lbs) | $P_{max}$ (PSI) | AIT (sec) |
|---|---|---|---|---|---|---|---|---|---|---|---|
| .357 Mag | Triton 125 JHP | 1409 | 25.16 | 10.09 | 10.09 | 2.70 | 2.70 | 48 | 551 | 1169 | 4.40 |
| 10mm | Magsafe96 | 1729 | 23.71 | 15.08 | 15.08 | 2.22 | 2.22 | 88 | 637 | 1014 | 4.48 |
| .40S&W | Magsafe84 | 1753 | 21.04 | 13.82 | 13.82 | 2.03 | 2.03 | 60 | 573 | 995 | 4.52 |
| .45ACP | Magsafe96 | 1644 | 22.55 | 14.13 | 14.13 | 1.85 | 1.85 | 76 | 576 | 1100 | 4.68 |
| .38Sp | Glaser 80 | 1667 | 19.05 | 7.85 | 7.85 | 4.50 | 4.50 | 25 | 494 | 1347 | 4.72 |
| .45ACP | Glaser140 | 1355 | 27.10 | 10.60 | 10.60 | 6.80 | 6.80 | 42 | 571 | 1454 | 4.72 |
| .38Sp | Magsafe65 | 1841 | 17.10 | 12.33 | 12.33 | 1.52 | 1.52 | 29 | 489 | 849 | 4.76 |
| .357 Mag | Glaser 80 | 1687 | 19.28 | 6.73 | 6.73 | 3.90 | 3.90 | 48 | 506 | 1609 | 4.82 |
| 9mm | Triton 115 JHP | 1301 | 21.37 | 8.92 | 8.92 | 2.40 | 2.40 | 55 | 432 | 1032 | 4.82 |
| .40S&W | Glaser105 | 1449 | 21.74 | 7.54 | 7.54 | 5.00 | 5.00 | 56 | 489 | 1558 | 5.34 |
| .357Mag | Rem125JHP | 1458 | 26.04 | 24.19 | 20.73 | 3.33 | 2.85 | 40 | 590 | 805 | 7.34 |
| 9mm | Glaser 80 | 1555 | 17.77 | 6.69 | 6.69 | 3.90 | 3.90 | 52 | 430 | 1367 | 7.42 |
| .357Mag | Fed125JHP | 1442 | 25.75 | 24.50 | 24.50 | 3.98 | 3.98 | 80 | 577 | 919 | 7.44 |
| .357Mag | Fed110JHP | 1351 | 21.23 | 12.56 | 15.08 | 1.26 | 1.51 | 60 | 446 | 852 | 7.72 |
| .357Mag | Win125JHP | 1382 | 24.68 | 25.25 | 22.62 | 3.79 | 3.39 | 44 | 530 | 756 | 7.76 |
| .357Mag | CCI125JHP | 1367 | 24.41 | 27.26 | 21.11 | 3.82 | 2.95 | 60 | 519 | 639 | 7.78 |
| .357Mag | Win145ST | 1285 | 26.62 | 29.20 | 24.50 | 4.73 | 3.98 | 34 | 532 | 710 | 7.86 |
| .40S&W | Win155ST | 1210 | 26.79 | 29.68 | 26.38 | 5.20 | 4.62 | 47 | 504 | 713 | 7.86 |
| .357Mag | Rem110JHP | 1334 | 20.96 | 17.98 | 17.98 | 2.37 | 2.38 | 34 | 435 | 769 | 7.90 |
| .40S&W | Fed155JHP | 1142 | 25.29 | 24.50 | 24.50 | 3.98 | 3.98 | 57 | 449 | 714 | 7.90 |
| 10mm | Win175ST | 1267 | 31.68 | 31.80 | 30.53 | 6.44 | 6.18 | 40 | 624 | 953 | 7.92 |
| .357Mag | Rem125SJHP | 1277 | 22.80 | 36.51 | 28.27 | 6.85 | 5.30 | 21 | 453 | 558 | 7.94 |
| .380ACP | Glaser 70 | 1313 | 13.13 | 5.58 | 5.58 | 2.80 | 2.80 | 15 | 268 | 1024 | 7.94 |
| .45ACP | Rem185JHP+P | 1124 | 29.71 | 21.25 | 20.73 | 2.91 | 2.85 | 58 | 519 | 806 | 7.98 |
| .357Mag | CCI140JHP | 1322 | 26.44 | 28.24 | 21.86 | 4.10 | 3.17 | 43 | 543 | 670 | 8.06 |
| 10mm | Fed180HS | 995 | 25.59 | 29.04 | 25.25 | 4.87 | 4.23 | 34 | 396 | 548 | 8.22 |
| .357Mag | Rem158SJHP | 1220 | 27.54 | 29.84 | 18.85 | 3.73 | 2.36 | 35 | 522 | 525 | 8.30 |
| .40S&W | Fed180HS | 991 | 25.48 | 35.34 | 28.27 | 6.63 | 5.30 | 39 | 393 | 500 | 8.32 |
| .40S&W | Rem155JHP | 1136 | 25.15 | 31.61 | 22.99 | 4.82 | 3.51 | 41 | 444 | 514 | 8.40 |
| .45ACP | Fed230HS | 847 | 27.83 | 29.40 | 29.40 | 5.73 | 5.73 | 28 | 366 | 583 | 8.40 |
| .357Mag | Fed158NY | 1188 | 26.81 | 32.13 | 23.37 | 4.98 | 3.62 | 25 | 495 | 573 | 8.42 |
| .40S&W | CB150JHP | 1183 | 25.35 | 24.19 | 20.73 | 3.33 | 2.85 | 48 | 466 | 636 | 8.42 |
| 10mmMV | Fed180JHP | 1018 | 26.18 | 32.04 | 25.63 | 5.45 | 4.36 | 33 | 414 | 527 | 8.46 |
| .45ACP | CB185JHP | 1156 | 30.55 | 24.85 | 24.85 | 4.35 | 4.35 | 29 | 549 | 928 | 8.56 |
| .40S&W | CB180JHP+P | 1044 | 26.85 | 35.34 | 28.27 | 6.63 | 5.30 | 39 | 436 | 555 | 8.66 |
| .45ACP | Win185ST | 1004 | 26.53 | 29.78 | 29.78 | 5.88 | 5.88 | 30 | 414 | 659 | 8.82 |
| .40S&W | Win180RSXT | 989 | 25.43 | 29.40 | 27.14 | 5.29 | 4.88 | 29 | 391 | 574 | 8.86 |
| 10mmMV | Rem180JHP | 996 | 25.61 | 29.21 | 23.37 | 4.53 | 3.62 | 30 | 396 | 505 | 8.88 |
| .40S&W | Rem180JHP | 988 | 25.41 | 34.49 | 22.99 | 5.26 | 3.51 | 32 | 390 | 414 | 8.90 |
| 9mm | Fed115JHP+P+ | 1311 | 21.54 | 20.73 | 20.73 | 2.85 | 2.85 | 45 | 439 | 699 | 8.90 |
| .45ACP | CCI200JHP | 936 | 26.74 | 23.92 | 23.92 | 4.84 | 4.84 | 25 | 389 | 791 | 8.92 |
| 9mm | CB115JHP | 1333 | 21.90 | 24.53 | 20.73 | 3.37 | 2.85 | 39 | 454 | 610 | 8.92 |
| 9mm | Fed124HS+P+ | 1267 | 22.44 | 28.20 | 25.25 | 4.72 | 4.23 | 45 | 442 | 630 | 8.96 |

| Cartridge | Load | V | MV | $A_{PCC}$ | $A_{PCC}^{12"}$ | $V_{PCC}$ | $V_{PCC}^{12"}$ | TSC | E | $P_{max}$ | AIT |
|---|---|---|---|---|---|---|---|---|---|---|---|
| | | (fps) | (lbs-sec) | (sq. in.) | (sq. in.) | (cu. in.) | (cu. in.) | (cu. in.) | (ft-lbs) | (PSI) | (sec) |
| .38Sp | CB115+P+ | 1243 | 20.42 | 28.06 | 21.86 | 4.07 | 3.17 | 27 | 395 | 489 | 8.98 |
| 9mm | Win115JHP+P+ | 1288 | 21.16 | 19.60 | 19.60 | 3.87 | 3.87 | 37 | 424 | 1024 | 8.98 |
| 9mm | Rem115JHP+P+ | 1290 | 21.19 | 23.38 | 22.99 | 3.57 | 3.51 | 37 | 425 | 665 | 8.98 |
| .45ACP | Win230RSXT | 829 | 27.24 | 31.85 | 29.40 | 6.21 | 5.73 | 25 | 351 | 516 | 9.14 |
| .45ACP | Fed185JHP | 1011 | 26.72 | 28.83 | 25.63 | 4.90 | 4.36 | 18 | 420 | 594 | 9.24 |
| 9mm | Fed124NYLHP | 1105 | 19.57 | 21.67 | 21.67 | 3.25 | 3.25 | 31 | 336 | 558 | 9.28 |
| 9mm | Fed124HS | 1126 | 19.95 | 25.25 | 22.62 | 3.79 | 3.39 | 36 | 349 | 498 | 9.28 |
| 9mm | Fed115JHP | 1175 | 19.30 | 20.42 | 20.42 | 3.32 | 3.32 | 27 | 353 | 673 | 9.30 |
| 9mm | Rem115JHP | 1166 | 19.16 | 28.24 | 23.37 | 4.38 | 3.62 | 19 | 347 | 457 | 9.36 |
| 9mm | Win115ST | 1199 | 19.70 | 18.09 | 18.09 | 3.26 | 3.26 | 17 | 367 | 876 | 9.36 |
| 9mm | Fed147HS | 958 | 20.12 | 28.75 | 24.12 | 4.58 | 3.86 | 31 | 300 | 400 | 9.58 |
| 9mm | Hor90JHP | 1286 | 16.53 | 19.22 | 19.22 | 3.27 | 3.27 | 24 | 330 | 701 | 9.62 |
| 9mm | Win147RSXT | 962 | 20.20 | 30.77 | 23.37 | 4.77 | 3.62 | 29 | 302 | 365 | 9.68 |
| 9mm | CCI115JHP | 1149 | 18.88 | 22.87 | 21.1 | 3.20 | 3.20 | 19 | 337 | 495 | 9.80 |
| 9mm | Fed147JHP9MS1 | 979 | 20.56 | 28.47 | 21.48 | 4.06 | 3.06 | 20 | 313 | 376 | 9.84 |
| 9mm | Win147JHP | 890 | 18.69 | 28.97 | 21.86 | 4.20 | 3.17 | 20 | 259 | 311 | 9.90 |
| .40S&W | Horn155XTP | 1157 | 25.62 | 29.90 | 25.63 | 5.08 | 4.36 | 45 | 461 | 629 | 10.38 |
| .38Sp | Win158LHP+P | 996 | 22.48 | 28.82 | 23.37 | 4.47 | 3.62 | 13 | 348 | 449 | 10.76 |
| .38Sp | Fed158LHP+P | 982 | 22.17 | 28.65 | 22.62 | 4.30 | 3.39 | 13 | 338 | 425 | 10.80 |
| .38Sp | Rem158LHP+P | 924 | 20.86 | 31.61 | 25.63 | 5.37 | 4.36 | 16 | 300 | 387 | 10.86 |
| .38Sp | Fed125JHP+P | 998 | 17.82 | 26.44 | 26.01 | 4.56 | 4.48 | 19 | 276 | 433 | 10.92 |
| .380ACP | Fed90HS | 1008 | 12.96 | 19.13 | 19.13 | 2.77 | 2.77 | 21 | 203 | 369 | 10.94 |
| .38Sp | Win110JHP+P+ | 1136 | 17.85 | 18.69 | 18.69 | 3.97 | 3.97 | 28 | 315 | 860 | 11.02 |
| .380ACP | FED90JHP | 1007 | 12.95 | 16.28 | 13.57 | 1.47 | 1.22 | 15 | 203 | 269 | 11.06 |
| .380ACP | CB90JHP+P | 1041 | 13.38 | 16.40 | 16.40 | 2.38 | 2.38 | 16 | 217 | 460 | 11.12 |
| .38Sp | CCI125JHP+P | 947 | 16.91 | 30.78 | 26.38 | 5.39 | 4.62 | 20 | 249 | 340 | 11.36 |
| .38Sp | Rem95SJHP+P | 1138 | 15.44 | 14.84 | 14.84 | 1.67 | 1.67 | 19 | 273 | 497 | 11.38 |
| .38Sp | Win110ST+P | 999 | 15.70 | 18.10 | 18.10 | 3.03 | 3.03 | 19 | 244 | 541 | 11.66 |
| .38Sp | Win125JHP+P | 938 | 16.75 | 27.15 | 25.25 | 4.55 | 4.23 | 19 | 244 | 362 | 11.70 |
| .38Sp | Rem125SJHP+P | 935 | 16.70 | 23.52 | 23.52 | 3.76 | 3.76 | 16 | 243 | 396 | 11.74 |
| .380ACP | Win85ST | 980 | 11.90 | 12.86 | 12.86 | 2.03 | 2.03 | 11 | 181 | 533 | 12.88 |
| .380ACP | CCI88JHP | 965 | 12.13 | 19.22 | 13.57 | 1.73 | 1.22 | 9 | 182 | 204 | 13.40 |
| .380ACP | Rem88JHP | 996 | 12.52 | 16.48 | 15.45 | 1.69 | 1.58 | 11 | 194 | 289 | 13.46 |
| 9mm | Win115FMJ | 1163 | 19.11 | 27.70 | 13.57 | 2.49 | 1.22 | 11 | 345 | 162 | 14.40 |
| .380ACP | Horn90XTP | 984 | 12.65 | 16.31 | 16.58 | 1.79 | 1.82 | 10 | 193 | 188 | 15.58 |
| .380ACP | Fed95FMJ | 934 | 12.68 | 19.22 | 13.57 | 1.73 | 1.22 | 9 | 184 | 124 | 22.80 |
| .38Sp | Fed158RNL | 708 | 15.98 | 31.66 | 13.57 | 2.85 | 1.22 | 10 | 176 | 72 | 33.68 |